# Controllable Synthesis of Submillimeter Ultrathin 2D Ferromagnetic $Cr_5Te_8$ Nanosheets by GaTe-Assisted CVD

*Hanxiang Wu[1,2], Jianfeng Guo[1,2,3], Hua Xu[1,2], Suonan Zhaxi[1,2], Shuo Mi[1,2], Le Wang[1,2], Shanshan Chen[1,2], Rui Xu[1,2], Wei Ji[1,2], Zhihai Cheng[1,2] and Fei Pang[1,2,\*]*

[1]Beijing Key Laboratory of Optoelectronic Functional Materials & Micro-nano Devices, School of Physics, Renmin University of China, Beijing 100872, China

[2]Key Laboratory of Quantum State Construction and Manipulation (Ministry of Education), Renmin University of China, Beijing 100872, China

[3]Institute of Physics, Chinese Academy of Sciences, Beijing 100190, China

*Email: feipang@ruc.edu.cn

**Abstract:** 2D non-van der Waals (vdW) $Cr_5Te_8$ has attracted widespread research interest for its air stability and thickness-dependent magnetic properties. However, exploring new methods for growing larger-scale ultrathin 2D $Cr_5Te_8$ remains challenging. Here, we selected GaTe powder as precursor to supply Te monomers and fabricated large-scale 2D $Cr_5Te_8$ nanosheets. By optimizing the growth temperature and source–substrate distance ($D_{SS}$), we successfully achieved $Cr_5Te_8$ nanosheets with lateral sizes of up to ~0.19 mm and corresponding thicknesses down to ~4.8 nm. The role of GaTe enhances the efficient Te atoms concentration, which promoted the lateral growth of $Cr_5Te_8$ nanosheets. Furthermore, our findings reveal the presence of $Cr_5Te_8$ nanosheets exhibiting serrated edges and a stacked structure like wedding cakes. Magnetic property measurement revealed the intense out-of-plane ferromagnetism in $Cr_5Te_8$, with the Curie temperature ($T_C$) of 172 K. This work paves a way for the controllable growth of the submillimeter ultrathin 2D ferromagnetic crystalline and lays the foundation for the future synthesis of millimeter ultrathin ferromagnets.

**Introduction**

Due to their significant potential for use in magnetic memory and spintronics,[1,2] 2D magnetic materials are attractive as an essential 2D materials member. These materials feature outstanding electrical,[3,4] photoelectrical,[5,6] and unique magnetic characteristics.[7-9] In particular, 2D non-van der Waals (vdW) chromium telluride ($Cr_xTe_y$) exhibits a wide range of thickness-dependent characteristics as well as novel magnetic properties for fundamental research and intriguing applications for their numerous self-intercalated phases.[10-22] Importantly, 2D $Cr_5Te_8$ nanosheets exhibit strong out-of-plane ferromagnetic properties, with the Curie temperature ($T_C$) increasing monotonically as the thickness increases. The strong interlayer coupling of the magnetic order plays a crucial role in the related phenomenon.[23-33] Moreover, magnetic bubbles and thickness-dependent maze-like magnetic domains were observed in $Cr_5Te_8$ nanosheets using cryogenic magnetic force microscopy (MFM).[23]

Compared to layered vdW materials, large-scale ultrathin covalent non-vdW materials are difficult to synthesis for their unsaturated dangling bonds on the surface. Until now, 2D $Cr_5Te_8$ crystals have been successfully synthesized with diverse sizes and thicknesses using Te powder via CVD techniques.[23-30] In previous works, the maximum lateral size of synthesized ultrathin $Cr_5Te_8$ crystals can reach around 20 μm, with corresponding thicknesses of approximately 6 to 7.8 nm.[24,27] The larger-scale $Cr_5Te_8$ with a maximum lateral size of approximately 200 μm was also fabricated, however, it is too thick up to 35.9 nm.[25] Thus, achieving ultrathin large-scale $Cr_5Te_8$ is still challenging.

Herein, we reported the controllable growth of submillimeter ultrathin 2D ferromagnetic $Cr_5Te_8$ with lateral size up to ~0.19 mm and thickness as low as ~4.8 nm via GaTe-assisted CVD. The GaTe as Te precursor can increase the effective Te atoms concentration by offering Te monomers, which is highlighted for the synthesis of ultrathin $Cr_5Te_8$. Meanwhile, we found submillimeter $Cr_5Te_8$ nanosheets show many morphological features such as serrated edges and stacked arrangements like wedding cakes. Furthermore, the growth behavior of 2D $Cr_5Te_8$ nanosheets was systematically studied by adjusting growth temperature and source–substrate distance ($D_{SS}$). Atomic force microscopy (AFM), X-ray diffraction (XRD), scanning electron microscopy (SEM) with energy dispersive spectroscopy (EDS), Raman and X-ray photoelectron spectroscopy (XPS) measurements illustrated the high crystallinity as well as accurate composition and structure of as-grown 2D $Cr_5Te_8$ nanosheets. Finally, magnetic property measurement system (MPMS) demonstrated that $Cr_5Te_8$ nanosheets possess ferromagnetism with the $T_C$ of 172 K.

## Material and methods

### GaTe-assisted CVD growth of $Cr_5Te_8$ nanosheets

The GaTe-assisted growth of 2D $Cr_5Te_8$ nanosheets were carried out in a single-zone tube furnace equipped with a 1-inch diameter quartz tube by an atmospheric pressure chemical vapor deposition (APCVD) method as shown in Figure 1a. GaTe (Alfa Aesar, purity 99.99%) and $CrCl_3$ (Alfa Aesar, purity 99.9%) powder were placed in the center of the furnace, where the temperature ranged from 600 to 800 °C. Then, freshly cleaved fluorophlogopite mica substrates ($KMg_3AlSi_3O_{10}F_2$) were used as the growth substrates, kept next to the powders of GaTe and $CrCl_3$. Prior to growth, the quartz tube was vacuumed and purged by Ar gas twice to remove the residue of oxygen and moisture. 100-sccm 5% $H_2$/Ar gas was used to transport the vapor species to the substrates and the growth time was 10 min.

**Transfer of $Cr_5Te_8$ nanosheets**

The as-grown $Cr_5Te_8$ nanosheets on mica were transferred to the target substrates ($SiO_2$/Si) using polystyrene (PS) as a medium for further characterization. Briefly, $Cr_5Te_8$ nanosheets grown on mica substrates were first spin-coated with PS solution at a speed of 3000 rpm for 1 min, then baked on a hot plate at 60 °C for 30 min to improve the adhesion between the nanosheets and PS. After the PS/$Cr_5Te_8$ was successfully transferred to the target substrate, it was baked at 80 °C for 20 min. Finally, the PS was removed with acetone.

**Characterization of 2D $Cr_5Te_8$**

The morphology and thickness of $Cr_5Te_8$ nanosheets were characterized by OM (6XB-PC, Shang Guang) and AFM (Dimension ICON, Bruker). The phase structure of the obtained nanosheets was analyzed by XRD (D8 ADVANCE, Bruker). The elemental composition and distribution were evaluated by SEM (NOVA NANOSEM450, FEI) equipped with EDS (X-MaxN 50 $mm^2$, Oxford Instruments). The vibration modes were ascertained with a confocal Raman microscopy (alpha300 R, WITec). The binding energy of $Cr_5Te_8$ nanosheets was analyzed by XPS (ESCALAB 250Xi, ThermoFisher Scientific).

The $T_C$ and magnetic hysteresis loop were performed in a magnetic property measurement system (MPMS3, Quantum Design), in which the anisotropic magnetic properties of the nanosheets grown on mica were observed. The temperature-dependent magnetic moments (*M-T*) for both out-of-plane ($H_\perp$) and in-plane ($H_{//}$) magnetic fields were measured within the temperature ranged from 1.8 to 400 K by the processes of zero-field cooling (ZFC) and field cooling (FC) with a field of 1 kOe. The field-dependent magnetizations (*M-H*) were carried out with the applied field ranged from −50 kOe to +50 kOe for $H_\perp$ and $H_{//}$ in several.

**Results and discussion**

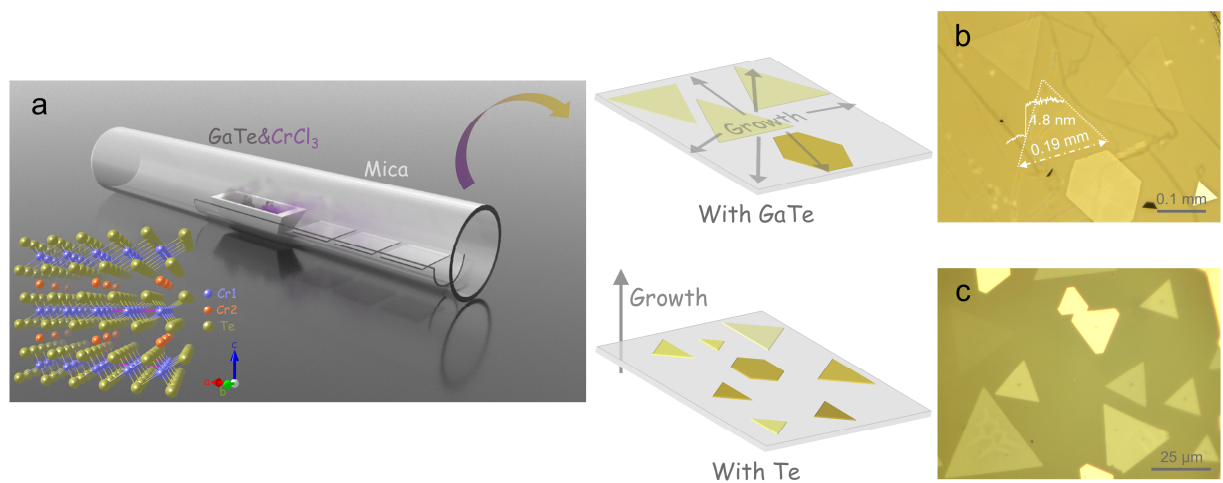

**Figure 1.** (a) Schematic diagram of GaTe-assisted CVD growth process of $Cr_5Te_8$ nanosheets. The inset shows the crystal structure (side view) of $Cr_5Te_8$. (b,c) Typical OM images of $Cr_5Te_8$ grown on mica substrates with GaTe (b) and Te (c) precursor.

To obtain 2D $Cr_5Te_8$, Te powder is commonly used as the Te source in traditional CVD growth techniques.[23-30] However, the Te powder can only supply few activated Te monomers suitable for chemical reactions. As a result, an insufficient effective Te concentration on the substrates gives rise to a Te rare environment unfavorable to growing 2D $Cr_5Te_8$.[34] Therefore, to address this issue, we design GaTe as a precursor to supply Te monomers for the growth of 2D $Cr_5Te_8$ nanosheets. The growth schematic diagram is described in Figure 1a. During the growth process of $Cr_5Te_8$, GaTe plays a facilitating role in enhancing its lateral growth. The results demonstrate 2D $Cr_5Te_8$ reached submillimeter levels by using GaTe precursor. The typical lateral size of $Cr_5Te_8$ nanosheets is up to 0.19 mm and its corresponding thickness lowers to 4.8 nm. The anisotropic ratio exceeds $10^4$. To the best of our knowledge, this represents the largest lateral size of ultrathin $Cr_5Te_8$ nanosheets achieved by CVD to date. In contrast, our previous work reported typical lateral sizes of ~30 μm for 2D $Cr_5Te_8$ grown using Te powder,[23] as shown in Figure 1b and 1c. The inset in Figure 1a shows the hexagonal crystal structure of $Cr_5Te_8$. The nonlayered crystal can be viewed as a self-intercalated structure in which 1/4 of the Cr atoms intercalate into the vdW gap between $CrTe_2$ layers.[32,33]

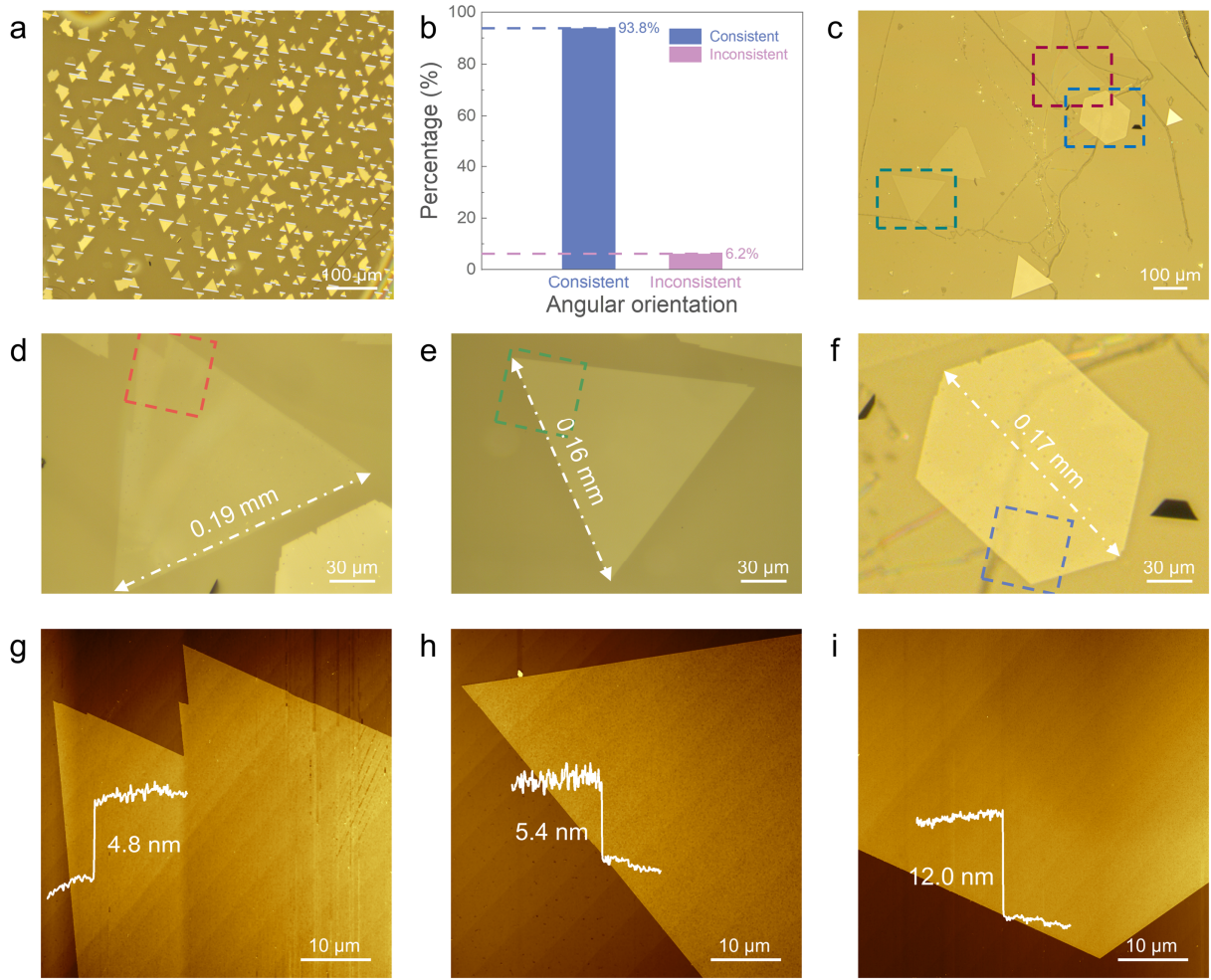

**Figure 2.** (a,b) Typical OM image (a) and corresponding statistical graph (b) of $Cr_5Te_8$ with high degree of orientation on mica. (c) Typical OM image of submillimeter $Cr_5Te_8$ on mica. (d-f) OM images of $Cr_5Te_8$ nanosheets within the red (d), green (e) and blue (f) box in (c). (g-i) AFM morphologies and corresponding height profiles within the red (g), green (h) and blue (i) box in (d-f).

Typical OM images of $Cr_5Te_8$ nanosheets under different growth conditions are shown in Figure 2a and 2c. The triangular nanosheets on mica exhibit a high degree of orientation in Figure 2a. According to the statistical data in Figure 2b, $Cr_5Te_8$ nanosheets with consistent angular orientation account for 93.8%. This high degree of orientation is attributed to an adequate and stable supply of Te atoms during the reaction. Figure 2c presents a typical OM image of submillimeter $Cr_5Te_8$. Figure 2d-2f respectively provide magnified views of the nanosheets within the red, green, and blue rectangular boxes in Figure 2c, with lateral sizes of 0.19 mm, 0.16 mm, and 0.17 mm. Figure 2g-2i display the AFM topography images and corresponding line profiles of the red, green, and blue boxed regions in Figure 2d-2f, revealing the corresponding nanosheet thicknesses of 4.8 nm, 5.2 nm, and 12.0 nm, respectively.

To investigate the crystal structure, the achieved nanosheets on mica were transferred to $SiO_2$/Si substrates and XRD characterization was carried out. As shown in Figure 3a, the four diffraction peaks located at 14.7°, 29.6°, 45.1° and 61.3° correspond to the (002), (004), (006), and (008) diffraction planes, which matches well with the standard PDF card of $Cr_5Te_8$ (PDF#50-1153), illustrating the accuracy of the as-grown 2D $Cr_5Te_8$ nanosheets structure.[32,33] Only the (00X) peaks appear in the XRD pattern, indicating that the as-grown 2D $Cr_5Te_8$ surface is parallel to the *ab* plane. In addition, the diffraction peak at 33.1° originates from the substrate, corresponding to the (200) crystal plane of Si. Further EDS was used to analyze the elemental composition of the as-grown $Cr_5Te_8$ nanosheets. As shown in Figure 3b, the atomic ratio of Cr and Te is approximately 5:8.03, consistent with the stoichiometric ratio of $Cr_5Te_8$. The absence of Ga in the EDS indicates that the as-grown nanosheet is free from any Ga impurities. EDS mapping of Cr, Te, and an overlay of them are shown in Figure 3c, 3d, and 3e, whose uniform color distribution manifests the uniformity of spatial distribution of Cr and Te, indicating the formation of $Cr_5Te_8$ crystal.

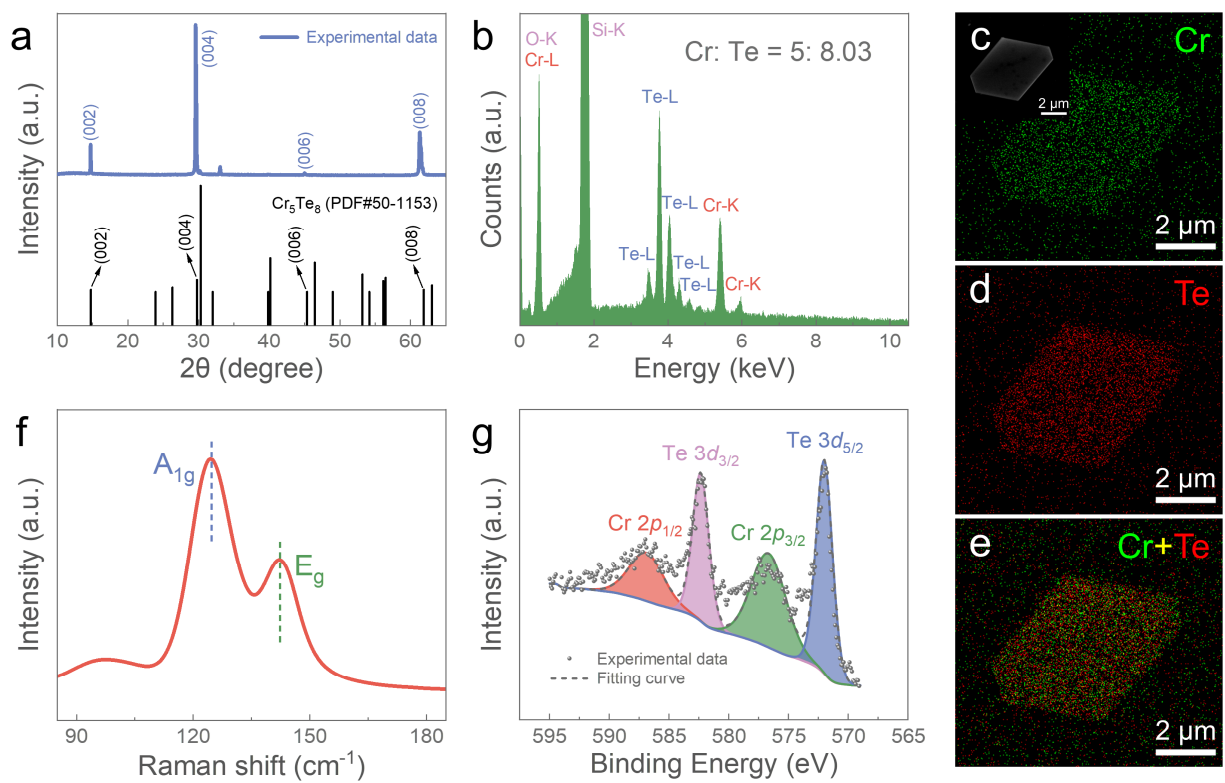

**Figure 3.** (a) XRD patterns of 2D $Cr_5Te_8$ nanosheets transferred to the $SiO_2$/Si substrate (top) and the standard PDF card (bottom). (b) EDS of as-grown 2D $Cr_5Te_8$ nanosheets. (c-e) EDS mapping of Cr element (c), Te element (d), and an overlay of Cr and Te elements (e) of a typical hexagonal $Cr_5Te_8$ nanosheet. The inset shows the corresponding SEM image. (f) Raman spectrum of an individual 2D $Cr_5Te_8$ nanosheet. (g) XPS of as-grown 2D $Cr_5Te_8$ nanosheets.

Furthermore, Raman spectrum was performed in an individual $Cr_5Te_8$ nanosheet, as shown in Figure 3f. Two prominent peaks at 125 $cm^{-1}$ and 142 $cm^{-1}$, corresponding to $A_{1g}$ and $E_g$ modes, appear for the $Cr_5Te_8$ nanosheet. No peak of GaTe was observed. Eventually, XPS was used to examine the chemical composition and bonding type of the as-grown $Cr_5Te_8$. As shown in Figure 3g, the peaks located at binding energies of ≈576.28 eV and 586.21 eV are attributed to Cr $2p_{3/2}$ and Cr $2p_{1/2}$, and the peaks located at ≈572.14 eV and 582.20 eV are attributed to Te $3d_{5/2}$ and Te $3d_{3/2}$, indicating a $Cr^{3.2+}$ state and $Te^{2-}$ state, well consistent with the valence state of $Cr_5Te_8$ crystal.[24] Besides, no discernible peaks of Ga element were observed.

Above all, these results confirm that 2D $Cr_5Te_8$ nanosheets with high quality and accurate composition have been successfully synthesized, providing an excellent material basis for further research on controllable growth and submillimeter synthesis.

To regulate the growth of $Cr_5Te_8$ nanosheets, detailed systematic studies were conducted and revealed that the growth temperature and $D_{SS}$ are essential for determining the average lateral size and nucleation density of $Cr_5Te_8$ nanosheets. To explore the effect of growth temperature, the other growth parameters (carrier gas and $D_{SS}$ = 7 cm) remained constant. Figure 4a-4d show typical OM images of $Cr_5Te_8$ grown on mica at different growth temperature. The average lateral sizes and nucleation densities of $Cr_5Te_8$ nanosheets is different noticeably with temperatures. When the growth temperature is 600 °C, $Cr_5Te_8$ nanosheets exhibit a tiny average lateral size of approximately 2.9 μm and a low nucleation density of around 3,450/mm². There is insufficient reactant since the growth temperature (600 °C) is lower than the evaporation temperature of the $CrCl_3$ and GaTe powder. Increasing to 700 °C, the average lateral size reaches roughly 6.4 μm, and the nucleation density increases substantially to 11,375/mm², indicating an ample supply of reactant at the appropriate growth temperature. Further increasing to 750 °C, the average lateral size continues to rise up to roughly 12.0 μm, but the nucleation density drops rapidly to 3,425/mm². At 800 °C, the average lateral size increases to an unusually large 50.0 μm, and the nucleation density reaches at a relatively tiny 196/mm². The lateral size and nucleation density of $Cr_5Te_8$ are influenced by the growth temperature, as illustrated in Figure 4i.

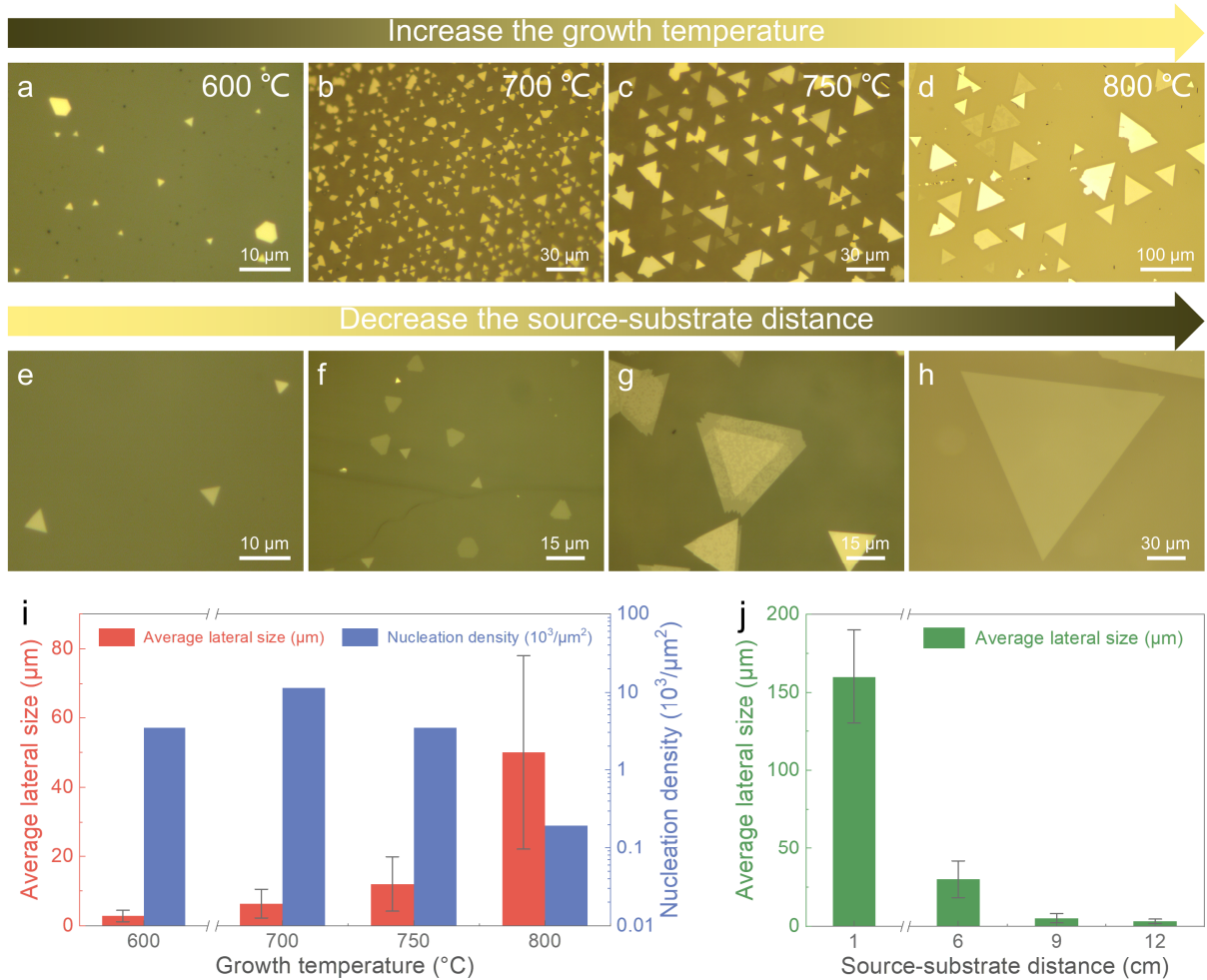

**Figure 4.** OM images of Cr$_5$Te$_8$ nanosheets synthesized at different growth temperature 600 °C (a), 700 °C (b), 750 °C (c), and 800 °C (d). OM images of Cr$_5$Te$_8$ nanosheets synthesized with different $D_{SS}$ of ≈12 cm (e), 9 cm (f), 6 cm (g), and 1 cm (h). (i) Histogram of average lateral size and nucleation density with variational growth temperature corresponding to (a-d). (j) Histogram of average lateral size with variational $D_{SS}$ corresponding to (e-h).

These growth regulatory laws are consistent with the nucleation model of vapor deposition proposed by W. K. Burton *et al.*, which demonstrated a positive correlation between growth temperature and lateral size but a negative correlation with nucleation likelihood.[35] The growth is predominantly governed by kinetics at lower growth temperature, which leads to a larger nucleation density and tinier nanosheets. The growth is mostly governed by thermodynamics at higher growth temperature, which lowers the nucleation density to produce bigger nanosheets.

We also discovered that $D_{SS}$ was another crucial parameter to control the growth of nanosheets. The $Cr_5Te_8$ nanosheets were synthesized with different $D_{SS}$ under the same growth conditions (carrier gas and growth temperature = 750 °C). As shown in Figure 4e–4h, the lateral sizes of $Cr_5Te_8$ nanosheets on mica varied with $D_{SS}$. According to the histogram shown in Figure 4j, the nanosheet size increased from 3.2 to 160 μm with decreasing $D_{SS}$ from 12 to 1 cm, because of the increasing effective Te concentrations.

We discovered a type of submillimeter ultrathin $Cr_5Te_8$ nanosheets with serrated edges, as depicted Figure 5a. The serrated $Cr_5Te_8$ nanosheet exhibits three smooth edges alternate with three serrated edges with a lateral size of ~157 μm, which was further characterized using AFM on a large scale, as shown in Figure 5b and 5c. Figure 5b displays the AFM morphology of the orange box area in Figure 5a, demonstrating that the serrated $Cr_5Te_8$ nanosheet possesses a consistent thickness of ~5 nm and a flat surface without impurities or faults beyond 80 × 80 μm$^2$ range. Figure 5c shows the AFM morphology of the green box area in Figure 5b, which confirms that the serrated edges angulate ~60°, consistent with the symmetry of $Cr_5Te_8$. These edges imply the high crystallinity of as-grown submillimeter ultrathin $Cr_5Te_8$ nanosheets.

In order to explore the growth mechanism of submillimeter ultrathin $Cr_5Te_8$, it is essential to comprehend the advantage of GaTe in lateral growth. In comparison with layered vdW materials, nonlayered $Cr_5Te_8$ forms via covalent bonding. When the nanosheets become thin to several nanometers, lots of unsaturated dangling bonds form on the surface, which will hinder 2D lateral growth. In our experiment, however, the GaTe precursor plays a crucial role. GaTe may supply the Te source in the form of active Te monomers, so they can react rapidly with metal precursors, obtaining a high lateral growth rate. The growth by active chalcogen monomer has been reported in other studies such as 2D metal dichalcogenides and their alloys.[34] In addition, the GaTe may act as the passivation agent, which prevented heterogeneous nucleation. The role of such passivators has been demonstrated in other studies such as $InCl_3$ and $In_2S_3$.[36,37] Therefore, we speculate that GaTe plays a key role in inhibiting the growth of nanosheets along the *c*-axis, leading to the successful growth of submillimeter ultrathin $Cr_5Te_8$ nanosheets.

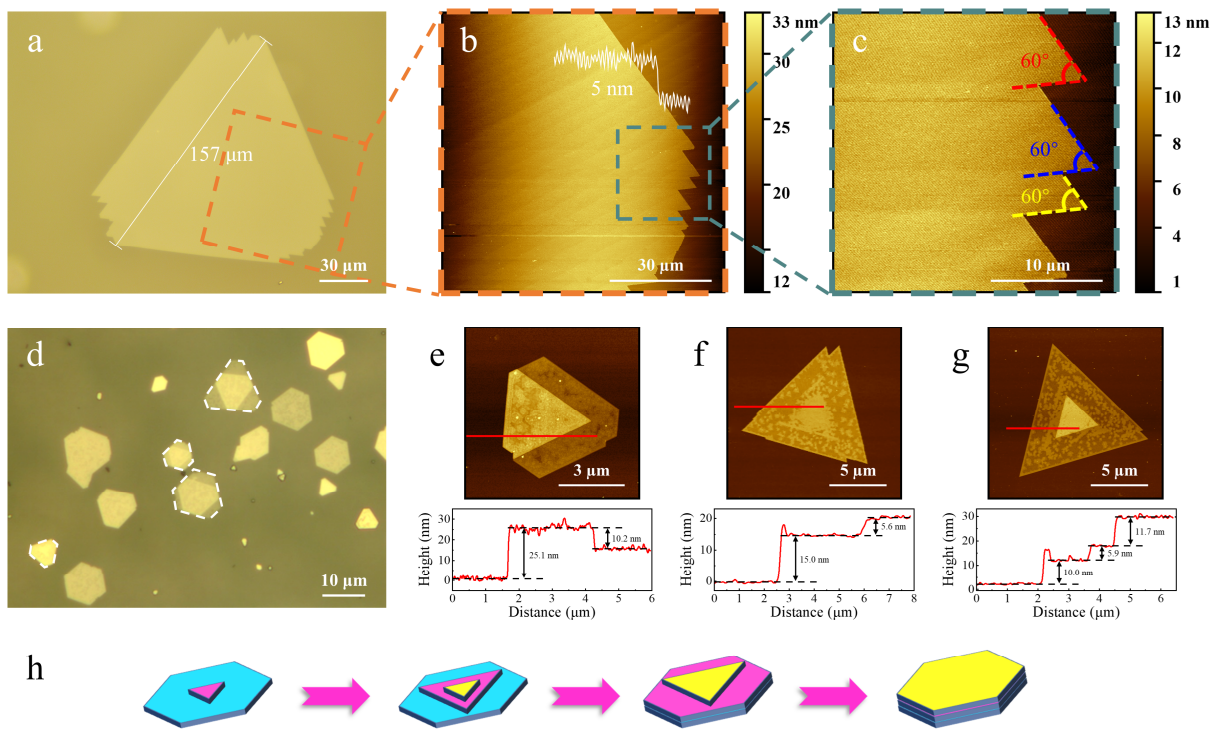

**Figure 5.** (a) Typical OM image of a serrated submillimeter ultrathin 2D $Cr_5Te_8$ nanosheet on mica. (b) AFM morphologies within the orange box in (a). (c) AFM morphologies within the green box in (b). (d) Typical OM image of stacked $Cr_5Te_8$ nanosheets grown on the mica substrate. (e-g) Typical AFM images and corresponding height profiles of $Cr_5Te_8$ nanosheets with different stacked arrangements reminiscent of wedding cakes. (h) Schematic diagram of "stacked growth" of 2D $Cr_5Te_8$ nanosheets.

Figure 5d shows a variety of typical morphologies of stacked $Cr_5Te_8$ nanosheets like wedding cakes grown on the mica substrate, which are composed of upper tiny hexagonal and triangular nanosheets on lower base nanosheets with or without a shared center of symmetry. The sublayer follows inner-to-outer growth modes. Multiple sublayer domains on $Cr_5Te_8$ nanosheets finally combine to form continuous layers. Figure 5e-5g show typical AFM images of $Cr_5Te_8$ with different stacked arrangements. The sublayers are not monolayers, and it is possible the unsaturated chemical bonds on surface of non-layered $Cr_5Te_8$ tend to adsorb external atoms to nucleation. So, we suppose the growth scenario of 2D stacked $Cr_5Te_8$. First, Cr and Te atoms are adsorbed on a complete $Cr_5Te_8$ nanosheet to nucleate. Then, as Cr and Te atoms accumulate, the nucleation site grows large to become a new upper sublayer on $Cr_5Te_8$ surface. Eventually, the sublayers grow, merge, and completely cover the whole $Cr_5Te_8$ crystal. The schematic diagram of the stacked growth of 2D $Cr_5Te_8$ nanosheets is described in Figure 5h.

MPMS was used to examine the magnetic characteristics of as-grown $Cr_5Te_8$ by GaTe-assisted CVD. The ZFC and FC curves with out-of-plane and in-plane under an external field of 1 kOe are shown in Figure 6a and 6d. The magnetic moment decreases with increasing temperature until it disappears around 172 K. This behavior demonstrates that the $Cr_5Te_8$ nanosheets have ferromagnetic order. Lower than those previously reported in bulk samples,[38] the first-order derivative of *M–T* in the inset shows that $Cr_5Te_8$ nanosheets have a magnetic phase transition from paramagnetism to ferromagnetism at $T_C$ = 172 K out of plane. More evidence that the *c*-axis is the magnetic easy axis and that ferromagnetic order has greater perpendicular magnetic anisotropy along *c*-axis is provided by the magnetization's strength parallel the *ab*-plane (*H*//*ab*) as compared to the perpendicular (*H*⊥*ab*).

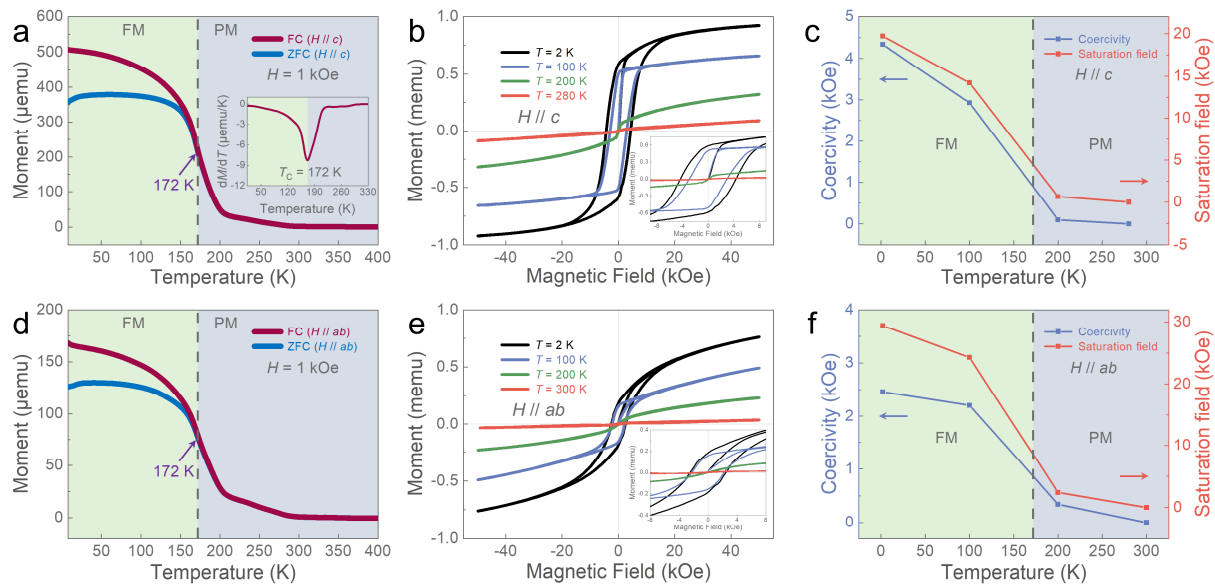

**Figure 6.** *M–T* with the magnetic field perpendicular (a) and parallel (d) to *ab*-plane on the sample for both FC (red point) process and ZFC (blue point) process under an external field of 1 kOe. The insets show the first-order derivatives of *M* with respect to *T* to derive the $T_C$ of $Cr_5Te_8$ (FC). *M–H* at different temperatures with the magnetic field perpendicular (b) and parallel (e) to the *ab*-plane on the nanosheets. Temperature-dependent coercivity and saturation field of $Cr_5Te_8$ nanosheets with the magnetic field perpendicular (c) and parallel (f) to the *ab*-plane on the nanosheets.

Further confirming the ferromagnetism and perpendicular magnetic anisotropy were the magnetic field dependent magnetic moments (*M–H*). Figure 6b and 6e show the *M–H* curves at various temperatures with *H*//*ab* and *H*⊥*ab*. The ferromagnetism of $Cr_5Te_8$ nanosheets is well demonstrated by the *M–H* hysteresis loops shown at 2 K and 100 K. Meanwhile, the ferromagnetism of $Cr_5Te_8$ is temperature-sensitive, as shown by the shrinking size of hysteresis loops with increasing temperature. Another proof that the *c*-axis is the magnetic easy axis came from larger magnetic hysteresis loops when *H*⊥*ab* and much smaller loops when *H*//*ab*.

Additionally, the temperature-dependent coercivity and saturation field indicate that $Cr_5Te_8$ nanosheets have a high coercive field up to 4337 Oe at 2K, as shown in Figure 6c and 6f. It infers that $Cr_5Te_8$ nanosheets can be potentially used in the hard magnetic application.

## Conclusions

In this paper, we successfully synthesized ultrathin 2D ferromagnetic $Cr_5Te_8$, whose lateral size is up to ~0.19 mm and corresponding thickness down to ~4.8 nm via GaTe-assisted CVD. This achievement signifies the highest lateral size attained so far for CVD-synthesized ultrathin $Cr_5Te_8$ nanosheets. Simultaneously, the growth temperature and $D_{SS}$ are key growth parameters to modulate the lateral size of $Cr_5Te_8$ nanosheets. Furthermore, the growth mechanism for 2D non-vdW $Cr_5Te_8$ involves stacked growth that varies in stacking sequences. Eventually, the as-synthesized $Cr_5Te_8$ nanosheets possess ferromagnetism with the $T_C$ of 172 K, which is lower than those previously reported in bulk samples. This research paves a way for the controlled synthesis of submillimeter ultrathin non-vdW 2D ferromagnetic crystals and may offer a fresh starting point for spintronics and magnetic memory devices applications.

## Acknowledgment

This project is supported by the National Natural Science Foundation of China (NSFC) (No. 12374200), the Ministry of Science and Technology (MOST) of China (No. 2023YFA1406500), the Strategic Priority Research Program (Chinese Academy of Sciences, CAS) (Grant No. XDB30000000), and the Fundamental Research Funds for the Central Universities and the Research Funds of Renmin University of China (No. 21XNLG27).